\documentclass[11pt,hidelinks]{article}

\usepackage{amsmath, amssymb, amsthm}
\usepackage{tikz}
\usepackage{mathpartir}
\usepackage[margin=0.5in]{geometry}
\usepackage{mathtools}
\usepackage{stmaryrd}

\newcommand{\p}{\textsf{p}}
\newcommand{\q}{\textsf{q}}
\newcommand{\C}{\textsf{whisk-L-R}}
\newcommand{\I}{\downdownarrows}
\newcommand{\J}{\rightrightarrows}
\newcommand{\concat}{\mathbin{\vcenter{\hbox{\scalebox{0.8}{$\blacksquare$}}}}}
\newcommand{\concatleft}{\concat\textsf{-1-L}}
\newcommand{\concatright}{\concat\textsf{-1-R}}
\newcommand{\concatrightnat}{\concat\textsf{-1-R-nat}}
\newcommand{\concatleftnat}{\concat\textsf{-1-L-nat}}
\newcommand{\Cleft}{\textsf{whisk-L-R-1-L}}
\newcommand{\Cright}{\textsf{whisk-L-R-1-R}}
\newcommand{\whiskerL}{\textsf{whisk-L}}
\newcommand{\whiskerR}{\textsf{whisk-R}}
\newcommand{\EH}{\textsf{EH}}
\newcommand{\EHleft}{\textsf{EH-1-L}}
\newcommand{\EHright}{\textsf{EH-1-R}}
\newcommand{\EHleftnat}{\textsf{EH-L-nat}}
\newcommand{\EHrightnat}{\textsf{EH-R-nat}}
\newcommand{\concatvert}{\boxminus}
\newcommand{\concathor}{\mathbin{\vcenter{\hbox{\scalebox{1.2}{$\boxbar$}}}}}
\newcommand{\invvert}{\upuparrows}
\newcommand{\invhor}{\leftleftarrows}
\newcommand{\ct}{%
  \mathchoice{\mathbin{\raisebox{0.5ex}{$\displaystyle\centerdot$}}}%
             {\mathbin{\raisebox{0.5ex}{$\centerdot$}}}%
             {\mathbin{\raisebox{0.25ex}{$\scriptstyle\,\centerdot\,$}}}%
             {\mathbin{\raisebox{0.1ex}{$\scriptscriptstyle\,\centerdot\,$}}}}
\newcommand{\ctt}{\ct\ct\,}
						
\newtheorem{lemma}{Lemma}
\newtheorem{theorem}{Theorem}

\begin{document}
\title{Syllepsis in Homotopy Type Theory}
\author{Kristina Sojakova}

\maketitle


\begin{abstract}
It is well-known that in homotopy type theory (HoTT)\cite{hott}, one can prove the \emph{Eckmann-Hilton} theorem: given two 2-loops $\p,\q : 1_\star = 1_\star$ on the reflexivity path at an arbitrary point $\star : A$, we have $\p \ct \q = \q \ct \p$. If we go one dimension higher, \emph{i.e.}, if $\p$ and $\q$ are 3-loops $\p,\q : 1_{1_\star} = 1_{1_\star}$, we show that a property classically known as \emph{syllepsis} also holds in HoTT: namely, the Eckmann-Hilton proof for $\q$ and $\p$ is the inverse of the Eckmann-Hilton proof for $\p$ and $\q$.
\end{abstract}


\section{The Eckmann-Hilton Argument}
Numerous short proofs of Eckmann-Hilton (EH) have been given by various people, including Favonia, Dan Christensen, and Mike Shulman. They are all propositionally equal but may have slightly different computational behavior. We start by outlining yet another such proof, which is particularly suitable for our proof of syllepsis. This proof is constructed differently from the aforementioned ones, but is again propositionally equal (it is in fact rather difficult to come up with a proof of EH that is \emph{not} propositionally equal to the existing ones). This will also serve to establish some notation and preliminary definitions.

If we go one dimension lower, EH will not hold: it is not hard to see that loops $p,q : a = a$ in general do not commute (for example, endofunctions in the universe). So to prove EH, we must make use of the fact that $\p$ and $\q$ are based at a path (and an identity one at that). One obvious thing we can do with a path is to compose it with another path; in this case, the only 1-path we have at our disposal is the reflexivity path $1_\star$ itself. Since all functions in HoTT are functorial, this gives us 2-paths $\whiskerL(1_\star,\p), \whiskerL(1_\star,\q) : 1_\star \ct 1_\star = 1_\star \ct 1_\star$ if we concatenate with $1_\star$ on the left and $\whiskerR(\p,1_\star), \whiskerR(\q,1_\star) : 1_\star \ct 1_\star = 1_\star \ct 1_\star$ if we concatenate with $1_\star$ on the right:

\begin{lemma}
For any points $a, b, c : A$, 1-paths $u : a = b$, $x, y : b = c$, and 2-path $q : x = y$, we have a term
\[ \whiskerL(u,p) : u \ct x = u \ct y\]   
Pictorially:
\begin{center}
\begin{tikzpicture}
\node (N0a) at (0,1) {$a$};
\node (N0b) at (2,1) {$b$};
\node (N0c) at (4,1) {$c$};
\node (N1a) at (0,0) {$a$};
\node (N1b) at (2,0) {$b$};
\node (N1c) at (4,0) {$c$};
\node at (3,0.5) {$q \Downarrow$};
\draw[-] (N0a) -- node[above]{$u$} (N0b);
\draw[-] (N0b) -- node[above]{$x$} (N0c);
\draw[-] (N1a) -- node[below]{$u$} (N1b);
\draw[-] (N1b) -- node[below]{$y$} (N1c);
\end{tikzpicture}
\end{center}
\end{lemma}

\begin{lemma}
For any points $a, b, c : A$, 1-paths $u, v : a = b$, $x : b = c$, and 2-path $p : u = v$, we have a term
\[ \whiskerR(q,x) : u \ct x = v \ct x \]   
Pictorially:
\begin{center}
\begin{tikzpicture}
\node (N0a) at (0,1) {$a$};
\node (N0b) at (2,1) {$b$};
\node (N0c) at (4,1) {$c$};
\node (N1a) at (0,0) {$a$};
\node (N1b) at (2,0) {$b$};
\node (N1c) at (4,0) {$c$};
\node at (1,0.5) {$p \Downarrow$};
\draw[-] (N0a) -- node[above]{$u$} (N0b);
\draw[-] (N0b) -- node[above]{$x$} (N0c);
\draw[-] (N1a) -- node[below]{$v$} (N1b);
\draw[-] (N1b) -- node[below]{$x$} (N1c);
\end{tikzpicture}
\end{center}
\end{lemma}

\noindent Whiskering is a special case of a \emph{parallel composition} of paths:

\begin{lemma}
For any points $a, b, c : A$, 1-loops $u, v : a = b$, $x, y : b = c$, and 2-paths $r : u = v$, $s : x = y$, we have a term
\[ p \ctt q : u \ct x = v \ct y \]   
Pictorially:
\begin{center}
\begin{tikzpicture}
\node (N0a) at (0,1) {$a$};
\node (N0b) at (2,1) {$b$};
\node (N0c) at (4,1) {$c$};
\node (N1a) at (0,0) {$a$};
\node (N1b) at (2,0) {$b$};
\node (N1c) at (4,0) {$c$};
\node at (1,0.5) {$p \Downarrow$};
\node at (3,0.5) {$q \Downarrow$};
\draw[-] (N0a) -- node[above]{$u$} (N0b);
\draw[-] (N0b) -- node[above]{$x$} (N0c);
\draw[-] (N1a) -- node[below]{$v$} (N1b);
\draw[-] (N1b) -- node[below]{$y$} (N1c);
\end{tikzpicture}
\end{center}
\end{lemma}

As the picture above suggests, parallel composition and ordinary composition of paths satisfy an exchange law: given 1-paths $w : a = b$, $z : b = c$ and 2-paths $r : v = w$, $s : y = z$, we have $(p \ctt q) \ct (r \ctt s) = (p \ct r) \ctt (q \ct s)$. It is not surprising, then, that the operations of whiskering on the left and whiskering on the right commute:

\begin{lemma}
For any points $a, b, c : A$, 1-paths $u, v : a = b$, $x, y : b = c$, and 2-paths $p : u = v$, $q : x = y$, we have a term
\[ \C(p,q) : \whiskerL(u,q) \ct \whiskerR(p,y) = \whiskerR(p,x) \ct \whiskerL(v,q) \]   
Pictorially:
\begin{center}
\begin{tikzpicture}
\node (N0) at (0,2) {$u \ct x$};
\node (N1) at (0,0) {$u \ct y$};
\node (N2) at (4,2) {$v \ct x$};
\node (N3) at (4,0) {$v \ct y$};
\draw[-] (N0) -- node[above]{$\whiskerR(p,x)$} (N2);
\draw[-] (N0) -- node[left]{$\whiskerL(u,q)$} (N1);
\draw[-] (N1) -- node[below]{$\whiskerR(p,y)$} (N3);
\draw[-] (N2) -- node[right]{$\whiskerL(v,q)$} (N3);
\end{tikzpicture}
\end{center}
\end{lemma}

\noindent In our case, this means the commuting diagram below:
\begin{center}
\begin{tikzpicture}
\node (N0) at (0,2) {$1_\star \ct 1_\star$};
\node (N1) at (0,0) {$1_\star \ct 1_\star$};
\node (N2) at (4,2) {$1_\star \ct 1_\star$};
\node (N3) at (4,0) {$1_\star \ct 1_\star$};
\draw[-] (N0) -- node[above]{\emph{$\whiskerR(\q,1_\star)$}} (N2);
\draw[-] (N0) -- node[left]{\emph{$\whiskerL(1_\star,\p)$}} (N1);
\draw[-] (N1) -- node[below]{\emph{$\whiskerR(\q,1_\star)$}} (N3);
\draw[-] (N2) -- node[right]{\emph{$\whiskerL(1_\star,\p)$}} (N3);
\end{tikzpicture}
\end{center}

On the other hand, $1_\star \ct 1_\star$ is just $1_\star$, so a reasonable question might be whether $\whiskerL(1_\star,\p)$ is equal to $\p$ and $\whiskerR(\q,1_\star)$ to $\q$. Of course, the answer is yes, so the above diagram essentially proves EH. Nevertheless, since we will be reasoning about EH computationally, we need to give the proof in detail.

To prove the equalities $\whiskerL(1_\star,\p) = \p$ and $\whiskerR(\q,1_\star) = \q$, we expand them into commuting diagrams:

\begin{lemma}
Concatenation on the left by reflexivity is natural: for any points $a, b : A$, 1-paths $u, v : a = b$, and 2-path $p : u = v$, we have a term
\[ \concatleftnat(p) : \whiskerL(1_a,p) \ct \concatleft(v) = \concatleft(u) \ct p \]   
Pictorially:
\begin{center}
\begin{tikzpicture}
\node (N0) at (0,2) {$1_a \ct u$};
\node (N1) at (0,0) {$1_a \ct v$};
\node (N2) at (4,2) {$u$};
\node (N3) at (4,0) {$v$};
\draw[-] (N0) -- node[above]{$\concatleft(u)$} (N2);
\draw[-] (N0) -- node[left]{$\whiskerL(1_a,p)$} (N1);
\draw[-] (N1) -- node[below]{$\concatleft(v)$} (N3);
\draw[-] (N2) -- node[right]{$p$} (N3);
\end{tikzpicture}
\end{center}
\end{lemma}

\begin{lemma}
Concatenation on the right by reflexivity is natural: for any points $b, c : A$, 1-paths $x, y : b = c$, and 2-path $q : x = y$, we have a term
\[ \concatrightnat(q) : \whiskerR(q,1_c) \ct \concatright(y) = \concatright(x) \ct q \]   
Pictorially:
\begin{center}
\begin{tikzpicture}
\node (N0) at (0,2) {$x \ct 1_c$};
\node (N1) at (0,0) {$y \ct 1_c$};
\node (N2) at (4,2) {$x$};
\node (N3) at (4,0) {$y$};
\draw[-] (N0) -- node[above]{$\concatright(x)$} (N2);
\draw[-] (N0) -- node[left]{$\whiskerR(q,1_c)$} (N1);
\draw[-] (N1) -- node[below]{$\concatright(y)$} (N3);
\draw[-] (N2) -- node[right]{$q$} (N3);
\end{tikzpicture}
\end{center}
\end{lemma}	

Instantiating to the case $\p, \q : 1_\star = 1_\star$, we get terms
\emph{
\begin{align*}
& \concatleftnat(\p) : \whiskerL(1_\star,\p) \ct 1_{1_\star} = 1_{1_\star} \ct \p \\
& \concatrightnat(\q) : \whiskerR(\q,1_\star) \ct 1_{1_\star} = 1_{1_\star} \ct \q
\end{align*}}%
This clearly gives us $\whiskerL(1_\star,\p) = \p$ and $\whiskerR(\q,1_\star) = \q$ as desired.

Squashing a commuting square whose opposing sides are reflexivities into an equality is so common that it will be useful to introduce a shorthand for this:

\begin{lemma}
For any points $a, b : A$ and paths $p, q : a = b$, we have an equivalence
\[ \I\, : (p \ct 1_b  = 1_\star \ct q) \simeq (p = q) \]   
\end{lemma}

\begin{lemma}
For any points $a, b : A$ and paths $p, q : a = b$, we have an equivalence
\[ \J\, : (1_\star \ct p = q \ct 1_b) \simeq (p = q) \]   
\end{lemma}

\noindent We can now formally prove Eckmann-Hilton:

\begin{theorem}[Eckmann-Hilton]
For any point $\star : A$ and 2-loops $\p, \q : 1_\star = 1_\star$, we have a 3-path
\[ \EH(\p,\q) : \p \ct \q  = \q \ct \p \]
defined as the composition of the following three paths:
\begin{center}
\begin{tikzpicture}
\node (N0) at (0,6) {$\p \ct \q$};
\node (N1) at (0,4) {$\whiskerL(1_\star,\p) \ct \whiskerR(\q,1_\star)$};
\node (N2) at (0,2) {$\whiskerR(\q,1_\star) \ct \whiskerL(1_\star,\p)$};
\node (N3) at (0,0) {$\q \ct \p$};
\draw[-] (N0) -- node[right]{\scriptsize $\Big(\I\big(\concatleftnat(\p)\big) \ct\ct \I\big(\concatrightnat(\q)\big)\Big)^{-1}$} (N1);
\draw[-] (N1) -- node[right]{\scriptsize $\C(\p,\q)$} (N2);
\draw[-] (N2) -- node[right]{\scriptsize $\I\big(\concatrightnat(\q)\big) \ct\ct \I\big(\concatleftnat(\p)\big)$} (N3);
\end{tikzpicture}
\end{center}
\end{theorem}


\section{Eckmann-Hilton on Reflexivity}
Since we want to prove something \emph{about} the Eckmann-Hilton proof, and we intend to use path induction to do so, it is unsurprising that we will need to know how the EH proof behaves on identity paths. When both loops are reflexivities, the Eckmann-Hilton proof is very simple: $\EH(1_{1_\star},1_{1_\star})$ automatically reduces to $1_{1_{1_\star}}$. If one of the loops is a reflexivity, say $\p$, there is an obvious candidate for $\EH(1_{1_\star},\q)$, namely
\begin{center}
\begin{tikzpicture}
\node (N0) at (0,0) {\emph{$1_{1_\star} \ct \q$}};
\node (N1) at (4,0) {\emph{$\q$}};
\node (N2) at (8,0) {\emph{$\q \ct 1_{1_\star}$}};
\draw[-] (N0) -- node[above]{\emph{$\concatleft(\q)$}} (N1);
\draw[-] (N1) -- node[above]{\emph{$\concatright(\q)^{-1}$}} (N2);
\end{tikzpicture}
\end{center}
It is tempting (and correct) to instead use the path $\J^{-1}(1_\q)$ here. However, this may or may not give us what we want, depending on how exactly we define the equality $\J$. In one obvious implementation, $\J^{-1}(1_\q)$ reduces to 
\emph{\[\big(\concatleft(\q) \ct 1_\q\big) \ct \concatright(\q)^{-1}\]}%
and this, while equal to the path above, has a gratuitous $1_\q$ term floating around. This would result in some extra work later on and extra work is best avoided.

So we want to construct a term
\emph{\[\EHleft(\q) : \EH(1_{1_\star}, \q) = \concatleft(\q) \ct \concatright(\q)^{-1}\]}%
We cannot of course use induction on $\q$ right away. But we can work towards generalizing the situation until we can. The term $\EH(1_{1_\star}, \q)$ reduces to the following composition:
\begin{center}
\begin{tikzpicture}
\node (N0) at (0,6) {\emph{$1_{1_\star} \ct \q$}};
\node (N1) at (0,4) {\emph{$1_{1_\star} \ct \whiskerR(\q,1_\star)$}};
\node (N2) at (0,2) {\emph{$\whiskerR(\q,1_\star) \ct 1_{1_\star}$}};
\node (N3) at (0,0) {\emph{$\q \ct 1_{1_\star}$}};
\draw[-] (N0) -- node[right]{\scriptsize \emph{$\Big(1_{1_{1_\star}} \ct\ct \I\big(\concatrightnat(\q)\big)\Big)^{-1}$}} (N1);
\draw[-] (N1) -- node[right]{\scriptsize \emph{$\C(1_{1_\star},\q)$}} (N2);
\draw[-] (N2) -- node[right]{\scriptsize \emph{$\I\big(\concatrightnat(\q)\big) \ctt 1_{1_{1_\star}}$}} (N3);
\end{tikzpicture}
\end{center}
Since $\C$ was defined by induction on both arguments, we first need to figure out what it does when only one of the arguments is reflexivity. Fortunately, this is very easy since we can just perform induction on the remaining argument:

\begin{lemma}
For any point $a : A$, 1-paths $u : a = b$, $x, y : b = c$, and 2-path $q : x = y$, we have a term
\[ \Cleft(q) : \C(1_u,q) = \big(\I^{-1}\big(1_{\whiskerL(u,q)}\big)\big)\]
where the two paths in question witness the commutativity of the diagram below:
\begin{center}
\begin{tikzpicture}
\node (N0) at (0,2) {$u \ct x$};
\node (N1) at (0,0) {$u \ct y$};
\node (N2) at (4,2) {$u \ct x$};
\node (N3) at (4,0) {$u \ct y$};
\draw[-] (N0) -- node[above]{$1_{u \ct x}$} (N2);
\draw[-] (N0) -- node[left]{$\whiskerL(u,q)$} (N1);
\draw[-] (N1) -- node[below]{$1_{u \ct y}$} (N3);
\draw[-] (N2) -- node[right]{$\whiskerL(u,q)$} (N3);
\end{tikzpicture}
\end{center}
\end{lemma}

\begin{lemma}
For any points $a, b, c : A$, 1-paths $u, v : a = b$, $x : b = c$, and 2-path $p : u = v$, we have a term
\[ \Cright(p) : \C(p,1_x) = \big(\J^{-1}\big(1_{\whiskerR(p,x)}\big)\big)\]
where the two paths in question witness the commutativity of the diagram below:
\begin{center}
\begin{tikzpicture}
\node (N0) at (0,2) {$u \ct x$};
\node (N1) at (0,0) {$u \ct x$};
\node (N2) at (4,2) {$v \ct x$};
\node (N3) at (4,0) {$v \ct x$};
\draw[-] (N0) -- node[above]{$\whiskerR(p,x)$} (N2);
\draw[-] (N0) -- node[left]{$1_{u \ct x}$} (N1);
\draw[-] (N1) -- node[below]{$\whiskerR(p,x)$} (N3);
\draw[-] (N2) -- node[right]{$1_{v \ct x}$} (N3);
\end{tikzpicture}
\end{center}
\end{lemma}

\noindent We can thus replace the middle segment $\C(1_{1_\star},\q)$ in $\EH(1_{1_\star}, \q)$ as follows:

\begin{center}
\begin{tikzpicture}
\node (N0) at (0,6) {\emph{$1_{1_\star} \ct \q$}};
\node (N1) at (0,4) {\emph{$1_{1_\star} \ct \whiskerR(\q,1_\star)$}};
\node (N2) at (0,2) {\emph{$\whiskerR(\q,1_\star) \ct 1_{1_\star}$}};
\node (N3) at (0,0) {\emph{$\q \ct 1_{1_\star}$}};
\draw[-] (N0) -- node[right]{\scriptsize \emph{$\Big(1_{1_{1_\star}} \ct\ct \I\big(\concatrightnat(\q)\big)\Big)^{-1}$}} (N1);
\draw[-] (N1) -- node[right]{\scriptsize \emph{$\J^{-1}\big(1_{\whiskerR(\q,1_\star)}\big)$}} (N2);
\draw[-] (N2) -- node[right]{\scriptsize \emph{$\I\big(\concatrightnat(\q)\big) \ctt 1_{1_{1_\star}}$}} (N3);
\end{tikzpicture}
\end{center}
Looking at the three segments, we see that it is possible to replace the diagram 
\emph{\[\concatrightnat(\q) : \whiskerR(\q,1_\star) \ct 1_{1_\star} = 1_{1_\star} \ct \q\]}%
with an abstract $\theta : \whiskerR(\q,1_\star) \ct 1_{1_\star} = 1_{1_\star} \ct \q$ of the same type:
\begin{center}
\begin{tikzpicture}
\node (N0) at (0,4.5) {\emph{$1_{1_\star} \ct \q$}};
\node (N1) at (0,3) {\emph{$1_{1_\star} \ct \whiskerR(\q,1_\star)$}};
\node (N2) at (0,1.5) {\emph{$\whiskerR(\q,1_\star) \ct 1_{1_\star}$}};
\node (N3) at (0,0) {\emph{$\q \ct 1_{1_\star}$}};
\draw[-] (N0) -- node[right]{\scriptsize \emph{$\big(1_{1_{1_\star}} \ct\ct \I(\theta)\big)^{-1}$}} (N1);
\draw[-] (N1) -- node[right]{\scriptsize \emph{$\J^{-1}\big(1_{\whiskerR(q,1_\star)}\big)$}} (N2);
\draw[-] (N2) -- node[right]{\scriptsize \emph{$\I(\theta) \ctt 1_{1_{1_\star}}$}} (N3);
\end{tikzpicture}
\end{center}
Having a path $\whiskerR(\q,1_\star) \ct 1_{1_\star} = 1_{1_\star} \ct \q$ is equivalent to having a path $\whiskerR(\q,1_\star) = \q$. If we managed to free the endpoints of the latter, we would be able to dispose of it by path induction. Fortunately, there is nothing that prevents us from doing so: we can just assume abstract loops $p, q : 1_\star = 1_\star$ in lieu of $\whiskerR(\q,1_\star)$ and $\q$, respectively. Our goal now becomes to prove that given $\theta : p \ct 1_{1_\star} = 1_{1_\star} \ct q$, the path
\begin{center}
\begin{tikzpicture}
\node (N0) at (0,4.5) {\emph{$1_{1_\star} \ct q$}};
\node (N1) at (0,3) {\emph{$1_{1_\star} \ct p$}};
\node (N2) at (0,1.5) {\emph{$p \ct 1_{1_\star}$}};
\node (N3) at (0,0) {\emph{$q \ct 1_{1_\star}$}};
\draw[-] (N0) -- node[right]{\scriptsize \emph{$\big(1_{1_{1_\star}} \ct\ct \I(\theta)\big)^{-1}$}} (N1);
\draw[-] (N1) -- node[right]{\scriptsize \emph{$\J^{-1}(1_p)$}} (N2);
\draw[-] (N2) -- node[right]{\scriptsize \emph{$\I(\theta) \ctt 1_{1_{1_\star}}$}} (N3);
\end{tikzpicture}
\end{center}
equals $\concatleft(q) \ct \concatright(q)^{-1}$. This is already looking good but we can do even better: there is no longer anything in our goal that would require $p$ and $q$ to be loops, or even 2-paths! So we can fully generalize our goal: given points $a, b : A$, 1-paths $p, q : a = b$, and a 2-path $\theta : p \ct 1_{1_\star} = 1_{1_\star} \ct q$, the path
\begin{center}
\begin{tikzpicture}
\node (N0) at (0,4.5) {$1_\star \ct q$};
\node (N1) at (0,3) {$1_\star \ct p$};
\node (N2) at (0,1.5) {$p \ct 1_b$};
\node (N3) at (0,0) {$q \ct 1_b$};
\draw[-] (N0) -- node[right]{\scriptsize $\big(1_{1_\star} \ct\ct \I(\theta)\big)^{-1}$} (N1);
\draw[-] (N1) -- node[right]{\scriptsize $\J^{-1}(1_p)$} (N2);
\draw[-] (N2) -- node[right]{\scriptsize $\I(\theta) \ctt 1_{1_b}$} (N3);
\end{tikzpicture}
\end{center}
equals $\concatleft(q) \ct \concatright(q)^{-1}$. We can now convert the type of $\theta$ to the equivalent $p = q$: given points $a, b : A$, 1-paths $p, q : a = b$, and a 2-path $\theta : p = q$, the path
\begin{center}
\begin{tikzpicture}
\node (N0) at (0,4.5) {$1_\star \ct q$};
\node (N1) at (0,3) {$1_\star \ct p$};
\node (N2) at (0,1.5) {$p \ct 1_b$};
\node (N3) at (0,0) {$q \ct 1_b$};
\draw[-] (N0) -- node[right]{\scriptsize $(1_{1_\star} \ctt \theta)^{-1}$} (N1);
\draw[-] (N1) -- node[right]{\scriptsize $\J^{-1}(1_p)$} (N2);
\draw[-] (N2) -- node[right]{\scriptsize $\theta \ctt 1_{1_b}$} (N3);
\end{tikzpicture}
\end{center}
equals $\concatleft(q) \ct \concatright(q)^{-1}$. But this is now trivial to prove: we first do induction on $\theta$, which collapses $p$ and $q$, and subsequently we do induction on $p$, which collapses $a$ and $b$, and reduces everything to reflexivity.

Analogously, we construct a term
\emph{\[\EHright(p) : \EH(p, 1_{1_\star}) = \concatright(p) \ct \concatleft(p)^{-1}\]}
Our verbose explanation notwithstanding, the entire argument in this section can be coded up in just a few lines of Coq code, the most complex of which are the theorem statements themselves. As a further important bonus of formalization, we get that $\EHleft(1_{1_\star})$ and $\EHright(1_{1_\star})$ reduce to $1_{1_{1_\star}}$. It is of course also easy, albeit less convincing, to verify this explicitly by carefully examining the construction we gave.


\section{Naturality of Eckmann-Hilton}
As with all constructions in homotopy type theory, the Eckmann-Hilton proof \emph{itself} respects equality:

\begin{lemma}
For any 2-loops $u, v, x : 1_\star = 1_\star$, and 3-path $q : u = v$, we have a term
\[ \EHleftnat(q,x) : \whiskerR(q,x) \ct \EH(v,x) = \EH(u,x) \ct \whiskerL(x,q) \]   
Pictorially:
\begin{center}
\begin{tikzpicture}
\node (N0) at (0,2) {$u \ct x$};
\node (N1) at (0,0) {$v \ct x$};
\node (N2) at (4,2) {$x \ct u$};
\node (N3) at (4,0) {$x \ct v$};
\draw[-] (N0) -- node[above]{$\EH(u,x)$} (N2);
\draw[-] (N0) -- node[left]{$\whiskerR(q,x)$} (N1);
\draw[-] (N1) -- node[below]{$\EH(v,x)$} (N3);
\draw[-] (N2) -- node[right]{$\whiskerL(x,q)$} (N3);
\end{tikzpicture}
\end{center}
\end{lemma}

\begin{lemma}
For any 2-loops $u, x, y : 1_\star = 1_\star$, and 3-path $p : x = y$, we have a term
\[ \EHrightnat(u,p) : \whiskerL(u,p) \ct \EH(u,y) = \EH(u,x) \ct \whiskerR(p,u) \]   
Pictorially:
\begin{center}
\begin{tikzpicture}
\node (N0) at (0,2) {$u \ct x$};
\node (N1) at (0,0) {$u \ct y$};
\node (N2) at (4,2) {$x \ct u$};
\node (N3) at (4,0) {$y \ct u$};
\draw[-] (N0) -- node[above]{$\EH(u,x)$} (N2);
\draw[-] (N0) -- node[left]{$\whiskerL(u,p)$} (N1);
\draw[-] (N1) -- node[below]{$\EH(u,y)$} (N3);
\draw[-] (N2) -- node[right]{$\whiskerR(p,u)$} (N3);
\end{tikzpicture}
\end{center}
\end{lemma}

In our case all 2-loops are reflexivities $1_{1_\star}$. The terms $\EHleftnat(\q,1_{1_\star})$ and $\EHrightnat(1_{1_\star},\p)$, however, are not going to compute directly, as $\p$ and $\q$ are nontrivial and even loops. As we now show, however, this is not a problem - it turns out that both $\EHleftnat(\q,1_{1_\star})$ and $\EHrightnat(1_{1_\star},\p)$ can be constructed explicitly by pasting together a few commutative squares! So why even bother with an inductive definition? The key is precisely the word \emph{inductive} - replacing the explicit construction we give below by the inductive definition we gave above effectively amounts to an abstraction that lies at the very heart of the entire proof of syllepsis.

To this end, we recall the following standard constructions on commutative squares:
 
\begin{lemma}
For any points $a, b, c, d, e, f : A$, 1-paths $p : a = b$, $q : b = c$, $r : d = e$, $s : e = f$, $u : a = d$, $v : b = e$, $w : c = f$, and 2-paths $\gamma : p \ct v = u \ct r$, $\delta : q \ct w = v \ct s$ as in the diagram
\begin{center}
\begin{tikzpicture}
\node (N0a) at (0,4) {$a$};
\node (N1a) at (0,2) {$b$};
\node (N2a) at (0,0) {$c$};
\node (N0b) at (2,4) {$d$};
\node (N1b) at (2,2) {$e$};
\node (N2b) at (2,0) {$f$};
\node at (1,1) {$\delta$};
\node at (1,3) {$\gamma$};
\draw[-] (N0a) -- node[left]{$p$} (N1a);
\draw[-] (N1a) -- node[left]{$q$} (N2a);
\draw[-] (N0b) -- node[right]{$r$} (N1b);
\draw[-] (N1b) -- node[right]{$s$} (N2b);
\draw[-] (N0a) -- node[above]{$u$} (N0b);
\draw[-] (N1a) -- node[above]{$v$} (N1b);
\draw[-] (N2a) -- node[below]{$w$} (N2b);
\end{tikzpicture}
\end{center}
we have a term
\[ \gamma \concatvert \delta : (p \ct q) \ct w = u \ct (r \ct s) \]   
\end{lemma}

\begin{lemma}
For any points $a, b, c, d, e, f : A$, 1-paths $p : a = b$, $q : b = c$, $r : d = e$, $s : e = f$, $u : a = d$, $v : b = e$, $w : c = f$, and 2-paths $\gamma : u \ct r = p \ct v$, $\delta : v \ct s = q \ct w$ as in the diagram
\begin{center}
\begin{tikzpicture}
\node (N0a) at (0,2) {$a$};
\node (N1a) at (2,2) {$b$};
\node (N2a) at (4,2) {$c$};
\node (N0b) at (0,0) {$d$};
\node (N1b) at (2,0) {$e$};
\node (N2b) at (4,0) {$f$};
\node at (1,1) {$\gamma$};
\node at (3,1) {$\delta$};
\draw[-] (N0a) -- node[above]{$p$} (N1a);
\draw[-] (N1a) -- node[above]{$q$} (N2a);
\draw[-] (N0b) -- node[below]{$r$} (N1b);
\draw[-] (N1b) -- node[below]{$s$} (N2b);
\draw[-] (N0a) -- node[left]{$u$} (N0b);
\draw[-] (N1a) -- node[left]{$v$} (N1b);
\draw[-] (N2a) -- node[right]{$w$} (N2b);
\end{tikzpicture}
\end{center}
we have a term
\[ \gamma \concathor \delta : u \ct (r \ct s) = (p \ct q) \ct w \]   
\end{lemma}

\begin{lemma}
For any points $a, b, c, d : A$, 1-paths $p : a = b$, $q : c = d$, $r : a = c$, $s : b = d$, and 2-path $\gamma : p \ct s = r \ct q$ as in the diagram
\begin{center}
\begin{tikzpicture}
\node (N0a) at (0,2) {$a$};
\node (N1a) at (2,2) {$c$};
\node (N0b) at (0,0) {$b$};
\node (N1b) at (2,0) {$d$};
\node at (1,1) {$\gamma$};
\draw[-] (N0a) -- node[above]{$r$} (N1a);
\draw[-] (N0a) -- node[left]{$p$} (N0b);
\draw[-] (N0b) -- node[below]{$s$} (N1b);
\draw[-] (N1a) -- node[right]{$q$} (N1b);
\end{tikzpicture}
\end{center}
we have a term
\begin{align*}
& \gamma^\invvert : p^{-1} \ct r = s \ct q^{-1}
\end{align*}
\end{lemma}

\noindent We also recall that the higher path $\EHleftnat(\q,1_{1_\star})$ witnesses the commutativity of the following square:
\begin{center}
\begin{tikzpicture}
\node (N0) at (0,2) {$1_{1_\star} \ct 1_{1_\star}$};
\node (N1) at (0,0) {$1_{1_\star} \ct 1_{1_\star}$};
\node (N2) at (4,2) {$1_{1_\star} \ct 1_{1_\star}$};
\node (N3) at (4,0) {$1_{1_\star} \ct 1_{1_\star}$};
\node at (2, 1) {\emph{$\EHleftnat(\q,1_{1_\star})$}};
\draw[-] (N0) -- node[above]{\emph{$\EH(1_{1_\star},1_{1_\star})$}} (N2);
\draw[-] (N0) -- node[left]{\emph{$\whiskerR(\q,1_{1_\star})$}} (N1);
\draw[-] (N1) -- node[below]{\emph{$\EH(1_{1_\star},1_{1_\star})$}} (N3);
\draw[-] (N2) -- node[right]{\emph{$\whiskerL(1_{1_\star},\q)$}} (N3);
\end{tikzpicture}
\end{center}
There is another way to fill this square - we horizontally compose $\concatrightnat(q)$ with the horizontal inverse of $\concatleftnat(q)$:
\begin{center}
\begin{tikzpicture}
\node (N0a) at (0,6) {$1_{1_\star} \ct 1_{1_\star}$};
\node (N1a) at (5,6) {$1_{1_\star}$};
\node (N2a) at (10,6) {$1_{1_\star} \ct 1_{1_\star}$};
\node (N0b) at (0,4) {$1_{1_\star} \ct 1_{1_\star}$};
\node (N1b) at (5,4) {$1_{1_\star}$};
\node (N2b) at (10,4) {$1_{1_\star} \ct 1_{1_\star}$};
\node at (2.5, 5) {\emph{$\concatrightnat(\q)$}};
\node at (7.5, 5) {\emph{$\concatleftnat(\q)^\invhor$}};
\draw[-] (N0a) -- node[above]{\emph{$\concatright(1_{1_\star})$}} (N1a);
\draw[-] (N1a) -- node[above]{\emph{$\concatleft(1_{1_\star})^{-1}$}} (N2a);
\draw[-] (N0b) -- node[below]{\emph{$\concatright(1_{1_\star})$}} (N1b);
\draw[-] (N1b) -- node[below]{\emph{$\concatleft(1_{1_\star})^{-1}$}} (N2b);
\draw[-] (N0a) -- node[left]{\emph{$\whiskerR(\q,1_{1_\star})$}} (N0b);
\draw[-] (N1a) -- node[left]{\emph{$\q$}} (N1b);
\draw[-] (N2a) -- node[right]{\emph{$\whiskerL(1_{1_\star},\q)$}} (N2b);
\end{tikzpicture}
\end{center}
This indeed works because $\EH(1_{1_\star},1_{1_\star})$, $\concatright(1_{1_\star})$, and $\concatleft(1_{1_\star})$ are all definitionally $1_{1_{1_\star}}$. We now show that the two higher paths filling the square are equal too:
\emph{\[\EHleftnat(\q,1_{1_\star}) = \concatrightnat(\q) \concathor \, \concatleftnat(\q)^\invhor\]}%
We would like to prove this by induction on $\q$; in fact we don't have much choice since $\EHleftnat(\q,1_{1_\star})$ is \emph{defined} by induction on $\q$. But, as mentioned above, $\q$ is a loop. So we need to free at least one of its endpoints. This is not so easy, however: the entire reason why the above equality even type-checks is that both endpoints of $\q$ are themselves reflexivities.

So we need to do some adjustment in our goal. Suppose we free the right endpoint, \emph{i.e.}, we replace $\q$ by an abstract $q : 1_{1_\star} = y$. We could also try to free both endpoints, of course, to get $q : x = y$ but the one-sided version has a distinct advantage: after performing induction on $q$, \emph{both} $q$ and $y$ reduce to reflexivity. The more reflexivities, the more things compute, so we go for this option.

Comparing the two sides of our goal, the left-hand side now witnesses the commutativity of the following square:

\begin{center}
\begin{tikzpicture}
\node (N0) at (0,2) {$1_{1_\star} \ct 1_{1_\star}$};
\node (N1) at (0,0) {$1_{1_\star} \ct y$};
\node (N2) at (4,2) {$1_{1_\star} \ct 1_{1_\star}$};
\node (N3) at (4,0) {$1_{1_\star} \ct y$};
\node at (2, 1) {\emph{$\EHleftnat(q,1_{1_\star})$}};
\draw[-] (N0) -- node[above]{\emph{$\EH(1_{1_\star},1_{1_\star})$}} (N2);
\draw[-] (N0) -- node[left]{\emph{$\whiskerR(q,1_{1_\star})$}} (N1);
\draw[-] (N1) -- node[below]{\emph{$\EH(y,1_{1_\star})$}} (N3);
\draw[-] (N2) -- node[right]{\emph{$\whiskerL(1_{1_\star},q)$}} (N3);
\end{tikzpicture}
\end{center}

\noindent On the right-hand side we have the following pasting of squares:

\begin{center}
\begin{tikzpicture}
\node (N0a) at (0,6) {$1_{1_\star} \ct 1_{1_\star}$};
\node (N1a) at (5,6) {$1_{1_\star}$};
\node (N2a) at (10,6) {$1_{1_\star} \ct 1_{1_\star}$};
\node (N0b) at (0,4) {$1_{1_\star} \ct 1_{1_\star}$};
\node (N1b) at (5,4) {$y$};
\node (N2b) at (10,4) {$y \ct 1_{1_\star}$};
\node at (2.5, 5) {\emph{$\concatrightnat(q)$}};
\node at (7.5, 5) {\emph{$\concatleftnat(q)^\invhor$}};
\draw[-] (N0a) -- node[above]{\emph{$\concatright(1_{1_\star})$}} (N1a);
\draw[-] (N1a) -- node[above]{\emph{$\concatleft(1_{1_\star})^{-1}$}} (N2a);
\draw[-] (N0b) -- node[below]{\emph{$\concatright(y)$}} (N1b);
\draw[-] (N1b) -- node[below]{\emph{$\concatleft(y)^{-1}$}} (N2b);
\draw[-] (N0a) -- node[left]{\emph{$\whiskerR(q,1_{1_\star})$}} (N0b);
\draw[-] (N1a) -- node[left]{\emph{$q$}} (N1b);
\draw[-] (N2a) -- node[right]{\emph{$\whiskerL(1_{1_\star},q)$}} (N2b);
\end{tikzpicture}
\end{center}

As we can see, the two diagrams are no longer identical - the former has $\EH(y,1_{1_\star})$ on the bottom and the latter has $\concatright(y) \ct \concatleft(y)^{-1}$ instead. But not all hope is lost! In the previous section, we constructed a higher path for precisely such an occasion, namely
\emph{\[\EHright(y) : \EH(y,1_{1_\star}) = \concatright(y) \ct \concatleft(y)^{-1}\]}%
So all we need to do is to adjust our goal as follows:
\emph{
\begin{align*}
\EHleftnat(q,1_{1_\star}) = \, & \whiskerL\big(\whiskerR(q,1_{1_\star}),\EHright(y)\big) \, \ct \\
                  & \big(\concatleftnat(q) \concathor \, \concatleftnat(q)^\invhor\big)
\end{align*}}%
This is now easy to show by induction on $q$. Specializing to the case of interest $\q : 1_{1_\star} = 1_{1_\star}$ yields the desired equality
 \emph{\[\EHleftnat(\q,1_{1_\star}) = \concatrightnat(\q) \concathor \, \concatleftnat(\q)^\invhor\]}%
since in this case the path $\EHright(1_{1_\star})$ reduces to $1_{1_{1_\star}}$, as remarked in the previous section. Entirely analogously, we have
 \emph{\[\EHrightnat(1_{1_\star},\p) = \concatleftnat(\p) \concathor \, \concatleftnat(\p)^\invhor\]}%


\section{Syllepsis}
We can now proceed with our proof of syllepsis. Formally, we wish to show the following:
\begin{theorem}[Syllepsis]
For any point $\star:A$ and 3-loops $\p,\q: 1_{1_\star} = 1_{1_\star}$, we have
\[ \EH(q,p) = \EH(q,p)^{-1}\]
Pictorially:
\begin{center}
\begin{tikzpicture}
\node (N0) at (4,6) {$\p \ct \q$};
\node (N1) at (0,4) {$\whiskerL(1_{1_\star},\p) \ct \whiskerR(\q,1_{1_\star})$};
\node (N2) at (0,2) {$\whiskerR(\q,1_{1_\star}) \ct \whiskerL(1_{1_\star},\p)$};
\node (N3) at (4,0) {$\q \ct \p$};
\node (N4) at (8,2) {$\whiskerL(1_{1_\star},\q) \ct \whiskerR(\p,1_{1_\star})$};
\node (N5) at (8,4) {$\whiskerR(\p,1_{1_\star}) \ct \whiskerL(1_{1_\star},\q)$};
\node at (-0.8,5.3) {\scriptsize $\Big(\I\big(\concatleftnat(\p)\big) \ct\ct \I\big(\concatrightnat(\q)\big)\Big)^{-1}$};
\node at (9,5.3) {\scriptsize $\Big(\I\big(\concatrightnat(\p)\big) \ct\ct \I\big(\concatleftnat(\q)\big)\Big)^{-1}$};
\node at (-0.8,0.8) {\scriptsize $\I\big(\concatrightnat(\q)\big) \ct\ct \I\big(\concatleftnat(\p)\big)$};
\node at (8.7,0.8) {\scriptsize $\I\big(\concatleftnat(\q)\big) \ct\ct \I\big(\concatrightnat(\p)\big)$};
\draw[->] (N0) -- node[above]{} (N1);
\draw[->] (N1) -- node[left]{\scriptsize $\C(\p,\q)$} (N2);
\draw[->] (N2) -- node[below]{} (N3);
\draw[->] (N4) -- node[below]{} (N3);
\draw[->] (N5) -- node[right]{\scriptsize $\C(\q,\p)^{-1}$} (N4);
\draw[->] (N0) -- node[above]{} (N5);
\end{tikzpicture}
\end{center}
\end{theorem}


\paragraph{Step 1}
The natural first step towards proving syllepsis is to split the hexagon into two triangles and a square:
\begin{center}
\begin{tikzpicture}
\node (N0) at (4,6) {\emph{$\p \ct \q$}};
\node (N1) at (0,4) {\emph{$\whiskerL(1_{1_\star},\p) \ct \whiskerR(\q,1_{1_\star})$}};
\node (N2) at (0,2) {\emph{$\whiskerR(\q,1_{1_\star}) \ct \whiskerL(1_{1_\star},\p)$}};
\node (N3) at (4,0) {\emph{$\q \ct \p$}};
\node (N4) at (8,2) {\emph{$\whiskerL(1_{1_\star},\q) \ct \whiskerR(\p,1_{1_\star})$}};
\node (N5) at (8,4) {\emph{$\whiskerR(\p,1_{1_\star}) \ct \whiskerL(1_{1_\star},\q)$}};
\node at (-0.8,5.3) {\scriptsize $\Big(\I\big(\concatleftnat(\p)\big) \ct\ct \I\big(\concatrightnat(\q)\big)\Big)^{-1}$};
\node at (9,5.3) {\scriptsize $\Big(\I\big(\concatrightnat(\p)\big) \ct\ct \I\big(\concatleftnat(\q)\big)\Big)^{-1}$};
\node at (-0.8,0.8) {\scriptsize $\I\big(\concatrightnat(\q)\big) \ct\ct \I\big(\concatleftnat(\p)\big)$};
\node at (8.7,0.8) {\scriptsize $\I\big(\concatleftnat(\q)\big) \ct\ct \I\big(\concatrightnat(\p)\big)$};
\draw[->] (N0) -- node[above]{} (N1);
\draw[->] (N1) -- node[left]{\scriptsize \emph{$\C(\p,\q)$}} (N2);
\draw[->] (N2) -- node[below]{} (N3);
\draw[->] (N4) -- node[below]{} (N3);
\draw[->] (N5) -- node[right]{\scriptsize \emph{$\C(\q,\p)^{-1}$}} (N4);
\draw[->] (N0) -- node[above]{} (N5);
\draw[->] (N1) -- node[above]{\scriptsize \emph{$??$}} (N5);
\draw[->] (N2) -- node[below]{\scriptsize \emph{$??$}} (N4);
\end{tikzpicture}
\end{center}
We would like to prove the commutativity of the square in the middle by induction on $\p$ and $\q$. We thus want to generalize the situation so that we have $p : x = y$ and $q : u = v$ for arbitrary 2-loops $x,y,u,v : 1_\star = 1_\star$. The obvious attempt at this leads us to the following:
\begin{center}
\begin{tikzpicture}
\node (N0) at (0,2) {\emph{$\whiskerL(u,p) \ct \whiskerR(q,y)$}};
\node (N1) at (0,0) {\emph{$\whiskerR(q,x) \ct \whiskerL(v,p)$}};
\node (N2) at (8,2) {\emph{$\whiskerR(p,u) \ct \whiskerL(y,q)$}};
\node (N3) at (8,0) {\emph{$\whiskerL(x,q) \ct \whiskerR(p,v)$}};
\draw[-] (N0) -- node[left]{\scriptsize \emph{$\C(p,q)$}} (N1);
\draw[-] (N2) -- node[right]{\scriptsize \emph{$\C(q,p)^{-1}$}} (N3);
\draw[-] (N0) -- node[above]{\scriptsize \emph{$??$}} (N2);
\draw[-] (N1) -- node[below]{\scriptsize \emph{$??$}} (N3);
\end{tikzpicture}
\end{center}
However, the vertices on the left are paths $u \ct x = v \ct y$ whereas those on the right are paths $x \ct u = y \ct v$. To make the endpoints align, we insert an Eckmann-Hilton proof on both sides. We are able to do this precisely because the $p$ and $q$ we started with were 3-loops; syllepsis does \emph{not} hold if we go one dimension lower (this observation is due to Jamie Vicary).

Our square now has the following form, where the vertical paths simply leave the EH term unchanged:

\begin{center}
\begin{tikzpicture}
\node (N0) at (0,2) {\emph{$\big(\whiskerL(u,p) \ct \whiskerR(q,y)\big) \ct \EH(v,y)$}};
\node (N1) at (0,0) {\emph{$\big(\whiskerR(q,x) \ct \whiskerL(v,p)\big) \ct \EH(v,y)$}};
\node (N2) at (9,2) {\emph{$\EH(u,x) \ct \big(\whiskerR(p,u) \ct \whiskerL(y,q)\big)$}};
\node (N3) at (9,0) {\emph{$\EH(u,x) \ct \big(\whiskerL(x,q) \ct \whiskerR(p,v)\big)$}};
\draw[-] (N0) -- node[left]{\scriptsize \emph{$\C(p,q) \ctt 1_{\EH(v,y)}$}} (N1);
\draw[-] (N2) -- node[right]{\scriptsize \emph{$1_{\EH(u,x)} \ctt \C(q,p)^{-1}$}} (N3);
\draw[-] (N0) -- node[above]{\scriptsize \emph{$??$}} (N2);
\draw[-] (N1) -- node[below]{\scriptsize \emph{$??$}} (N3);
\end{tikzpicture}
\end{center}

Before we fill the horizontal paths and the square itself, we check that we have not lost touch with what we originally set out to prove: in the special case when all of the 2-paths are reflexivities, the EH terms $\EH(u,x)$ and $\EH(v,y)$ reduce to $1_{1_{1_\star}}$, as observed in the previous section. So the square we have is not \emph{exactly} the one we wanted - all the vertices now contain an extra reflexivity path -- but it is close enough.

To construct the horizontal paths, we need to fill the following two diagrams:
\begin{center}
\begin{tikzpicture}
\node (N0a) at (0,4) {$u \ct x$};
\node (N1a) at (0,2) {$u \ct y$};
\node (N2a) at (0,0) {$v \ct y$};
\node (N0b) at (3,4) {$x \ct u$};
\node (N1b) at (3,2) {$y \ct u$};
\node (N2b) at (3,0) {$y \ct v$};
\node (N0c) at (9,4) {$u \ct x$};
\node (N1c) at (9,2) {$v \ct x$};
\node (N2c) at (9,0) {$v \ct y$};
\node (N0d) at (12,4) {$x \ct u$};
\node (N1d) at (12,2) {$x \ct v$};
\node (N2d) at (12,0) {$y \ct v$};
\draw[-] (N0a) -- node[left]{\emph{$\whiskerL(u,p)$}} (N1a);
\draw[-] (N1a) -- node[left]{\emph{$\whiskerR(q,y)$}} (N2a);
\draw[-] (N0b) -- node[right]{\emph{$\whiskerR(p,u)$}} (N1b);
\draw[-] (N1b) -- node[right]{\emph{$\whiskerL(y,q)$}} (N2b);
\draw[-] (N0c) -- node[left]{\emph{$\whiskerR(q,x)$}} (N1c);
\draw[-] (N1c) -- node[left]{\emph{$\whiskerL(v,p)$}} (N2c);
\draw[-] (N0d) -- node[right]{\emph{$\whiskerL(x,q)$}} (N1d);
\draw[-] (N1d) -- node[right]{\emph{$\whiskerR(p,v)$}} (N2d);
\draw[-] (N0a) -- node[above]{\emph{$\EH(u,x)$}} (N0b);
\draw[-] (N2a) -- node[below]{\emph{$\EH(v,y)$}} (N2b);
\draw[-] (N0c) -- node[above]{\emph{$\EH(u,x)$}} (N0d);
\draw[-] (N2c) -- node[below]{\emph{$\EH(v,y)$}} (N2d);
\end{tikzpicture}
\end{center}
The obvious way to do this is to split each diagram into two squares as follows:
\begin{center}
\begin{tikzpicture}
\node (N0a) at (0,4) {$u \ct x$};
\node (N1a) at (0,2) {$u \ct y$};
\node (N2a) at (0,0) {$v \ct y$};
\node (N0b) at (3,4) {$x \ct u$};
\node (N1b) at (3,2) {$y \ct u$};
\node (N2b) at (3,0) {$y \ct v$};
\node (N0c) at (9,4) {$u \ct x$};
\node (N1c) at (9,2) {$v \ct x$};
\node (N2c) at (9,0) {$v \ct y$};
\node (N0d) at (12,4) {$x \ct u$};
\node (N1d) at (12,2) {$x \ct v$};
\node (N2d) at (12,0) {$y \ct v$};
\draw[-] (N0a) -- node[left]{\emph{$\whiskerL(u,p)$}} (N1a);
\draw[-] (N1a) -- node[left]{\emph{$\whiskerR(q,y)$}} (N2a);
\draw[-] (N0b) -- node[right]{\emph{$\whiskerR(p,u)$}} (N1b);
\draw[-] (N1b) -- node[right]{\emph{$\whiskerL(y,q)$}} (N2b);
\draw[-] (N0c) -- node[left]{\emph{$\whiskerR(q,x)$}} (N1c);
\draw[-] (N1c) -- node[left]{\emph{$\whiskerL(v,p)$}} (N2c);
\draw[-] (N0d) -- node[right]{\emph{$\whiskerL(x,q)$}} (N1d);
\draw[-] (N1d) -- node[right]{\emph{$\whiskerR(p,v)$}} (N2d);
\draw[-] (N0a) -- node[above]{\emph{$\EH(u,x)$}} (N0b);
\draw[-] (N1a) -- node[above]{\emph{$\EH(u,y)$}} (N1b);
\draw[-] (N2a) -- node[below]{\emph{$\EH(v,y)$}} (N2b);
\draw[-] (N0c) -- node[above]{\emph{$\EH(u,x)$}} (N0d);
\draw[-] (N1c) -- node[above]{\emph{$\EH(v,x)$}} (N1d);
\draw[-] (N2c) -- node[below]{\emph{$\EH(v,y)$}} (N2d);
\end{tikzpicture}
\end{center}
Each small square now commutes by the naturality of Eckmann-Hilton, so we can fill the horizontal paths in our big square as follows:
\begin{center}
\begin{tikzpicture}
\node (N0) at (0,2) {\emph{$\big(\whiskerL(u,p) \ct \whiskerR(q,y)\big) \ct \EH(v,y)$}};
\node (N1) at (0,0) {\emph{$\big(\whiskerR(q,x) \ct \whiskerL(v,p)\big) \ct \EH(v,y)$}};
\node (N2) at (11,2) {\emph{$\EH(u,x) \ct \big(\whiskerR(p,u) \ct \whiskerL(y,q)\big)$}};
\node (N3) at (11,0) {\emph{$\EH(u,x) \ct \big(\whiskerL(x,q) \ct \whiskerR(p,v)\big)$}};
\draw[-] (N0) -- node[left]{\scriptsize \emph{$\C(p,q) \ctt 1_{\EH(v,y)}$}} (N1);
\draw[-] (N2) -- node[right]{\scriptsize \emph{$1_{\EH(u,x)} \ctt \C(q,p)^{-1}$}} (N3);
\draw[-] (N0) -- node[above]{\scriptsize \emph{$\EHrightnat(u,p) \concatvert \EHleftnat(q,y)$}} (N2);
\draw[-] (N1) -- node[below]{\scriptsize \emph{$\EHleftnat(q,x) \concatvert \EHrightnat(v,p)$}} (N3);
\end{tikzpicture}
\end{center}
The above commutes by induction on $p$ and $q$. Specializing the generalization $p : x = y$ and $q : u = v$ to our situation yields the following commuting square:

\begin{center}
\begin{tikzpicture}
\node (N0) at (0,2) {\emph{$\big(\whiskerL(1_{1_\star},\p) \ct \whiskerR(\q,1_{1_\star})\big) \ct 1_{1_{1_\star}}$}};
\node (N1) at (0,0) {\emph{$\big(\whiskerR(\q,1_{1_\star}) \ct \whiskerL(1_{1_\star},\p)\big) \ct 1_{1_{1_\star}}$}};
\node (N2) at (11.5,2) {\emph{$1_{1_{1_\star}} \ct \big(\whiskerR(\p,1_{1_\star}) \ct \whiskerL(1_{1_\star},\q)\big)$}};
\node (N3) at (11.5,0) {\emph{$1_{1_{1_\star}} \ct \big(\whiskerL(1_{1_\star},\q) \ct \whiskerR(\p,1_{1_\star})\big)$}};
\draw[-] (N0) -- node[left]{\scriptsize \emph{$\C(\p,\q) \ctt 1_{1_{1_{1_\star}}}$}} (N1);
\draw[-] (N2) -- node[right]{\scriptsize \emph{$1_{1_{1_{1_\star}}} \ctt \C(\q,\p)^{-1}$}} (N3);
\draw[-] (N0) -- node[above]{\scriptsize \emph{$\EHrightnat(1_{1_\star},\p) \concatvert \EHleftnat(\q,1_{1_\star})$}} (N2);
\draw[-] (N1) -- node[below]{\scriptsize \emph{$\EHleftnat(\q,1_{1_\star}) \concatvert \EHrightnat(1_{1_\star},\p)$}} (N3);
\end{tikzpicture}
\end{center}


\paragraph{Step 2}
We now want to somehow retrofit the square from the previous step into the original diagram for syllepsis. Of course we cannot do this quite literally, since our square has an extra reflexivity path in each vertex. But we can fit the following triangles into the upper and lower part of the hexagon, respectively:
\begin{center}
\begin{tikzpicture}
\node (N0a) at (0,4.5) {\emph{$\whiskerL(1_{1_\star},\p) \ct \whiskerR(\q,1_{1_\star})$}};
\node (N1a) at (5.5,7) {\emph{$\p \ct \q$}};
\node (N2a) at (11,4.5) {\emph{$\whiskerR(\p,1_{1_\star}) \ct \whiskerL(1_{1_\star},\q)$}};
\node (N0b) at (0,2.5) {\emph{$\whiskerR(\q,1_{1_\star}) \ct \whiskerL(1_{1_\star},\p)$}};
\node (N1b) at (5.5,0) {\emph{$\q \ct \p$}};
\node (N2b) at (11,2.5) {\emph{$\whiskerL(1_{1_\star},q) \ct \whiskerR(p,1_{1_\star})$}};
\node at (-0.2,5.9) {\scriptsize \emph{$\Big(\I\big(\concatleftnat(\p)\big) \ct\ct \I\big(\concatrightnat(\q)\big)\Big)^{-1}$}};
\node at (11.5,5.9) {\scriptsize \emph{$\Big(\I\big(\concatrightnat(\p)\big) \ct\ct \I\big(\concatleftnat(\q)\big)\Big)^{-1}$}};
\node at (0.3,0.9) {\scriptsize \emph{$\I\big(\concatrightnat(\q)\big) \ct\ct \I\big(\concatleftnat(\p)\big)$}};
\node at (10.5,0.9) {\scriptsize \emph{$\I\big(\concatleftnat(\q)\big) \ct\ct \I\big(\concatrightnat(\p)\big)$}};
\draw[->] (N1a) -- node[left]{} (N0a);
\draw[->] (N1a) -- node[right]{} (N2a);
\draw[->] (N0a) -- node[below]{\scriptsize \emph{$\I\Big(\EHrightnat(1_{1_\star},\p) \concatvert \EHleftnat(\q,1_{1_\star})\Big)$}} (N2a);
\draw[->] (N0b) -- node[left]{} (N1b);
\draw[->] (N1b) -- node[right]{} (N2b);
\draw[->] (N0b) -- node[above]{\scriptsize \emph{$\I\Big(\EHleftnat(\q,1_{1_\star}) \concatvert \EHrightnat(1_{1_\star},\p)\Big)$}} (N2b);
\end{tikzpicture}
\end{center}

If we show that these triangles do in fact commute, we will be (almost) done. To do so, we again wish to suitably generalize the situation. Starting with the first triangle, we can put together the four commuting squares $\concatleftnat(\p), \concatrightnat(\p), \concatleftnat(\q), \concatrightnat(\q)$ as follows:

\begin{center}
\begin{tikzpicture}
\node (N0a) at (0,6) {$1_{1_\star} \ct 1_{1_\star}$};
\node (N1a) at (4,6) {$1_{1_\star}$};
\node (N2a) at (8,6) {$1_{1_\star} \ct 1_{1_\star}$};
\node (N0b) at (0,4) {$1_{1_\star} \ct 1_{1_\star}$};
\node (N1b) at (4,4) {$1_{1_\star}$};
\node (N2b) at (8,4) {$1_{1_\star} \ct 1_{1_\star}$};
\node (N0c) at (0,2) {$1_{1_\star} \ct 1_{1_\star}$};
\node (N1c) at (4,2) {$1_{1_\star}$};
\node (N2c) at (8,2) {$1_{1_\star} \ct 1_{1_\star}$};
\node at (2, 5) {\emph{$\concatleftnat(\p)$}};
\node at (6, 5) {\emph{$\big(\concatrightnat(\p)\big)^\invhor$}};
\node at (2, 3) {\emph{$\concatrightnat(\q)$}};
\node at (6, 3) {\emph{$\big(\concatleftnat(\q)\big)^\invhor$}};
\draw[-] (N0a) -- node[above]{$1_{1_{1_\star}}$} (N1a);
\draw[-] (N1a) -- node[above]{$1_{1_{1_\star}}$} (N2a);
\draw[-] (N0b) -- node[above]{$1_{1_{1_\star}}$} (N1b);
\draw[-] (N1b) -- node[above]{$1_{1_{1_\star}}$} (N2b);
\draw[-] (N0c) -- node[below]{$1_{1_{1_\star}}$} (N1c);
\draw[-] (N1c) -- node[below]{$1_{1_{1_\star}}$} (N2c);
\draw[-] (N0a) -- node[left]{\emph{$\whiskerL(1_{1_\star},\p)$}} (N0b);
\draw[-] (N0b) -- node[left]{\emph{$\whiskerR(\q,1_{1_\star})$}} (N0c);
\draw[-] (N1a) -- node[left]{\emph{$\p$}} (N1b);
\draw[-] (N1b) -- node[left]{\emph{$\q$}} (N1c);
\draw[-] (N2a) -- node[right]{\emph{$\whiskerR(\p,1_{1_\star})$}} (N2b);
\draw[-] (N2b) -- node[right]{\emph{$\whiskerL(1_{1_\star},\q)$}} (N2c);
\end{tikzpicture}
\end{center}
On the other hand, we also have the two commuting squares $\EHleftnat(\q,1_{1_\star})$ and $\EHrightnat(1_{1_\star},\p)$:
\begin{center}
\begin{tikzpicture}
\node (N0a) at (0,6) {$1_{1_\star} \ct 1_{1_\star}$};
\node (N1a) at (4,6) {$1_{1_\star}$};
\node (N2a) at (8,6) {$1_{1_\star} \ct 1_{1_\star}$};
\node (N0b) at (0,4) {$1_{1_\star} \ct 1_{1_\star}$};
\node (N1b) at (4,4) {$1_{1_\star}$};
\node (N2b) at (8,4) {$1_{1_\star} \ct 1_{1_\star}$};
\node (N0c) at (0,2) {$1_{1_\star} \ct 1_{1_\star}$};
\node (N1c) at (4,2) {$1_{1_\star}$};
\node (N2c) at (8,2) {$1_{1_\star} \ct 1_{1_\star}$};
\node at (4, 5) {\emph{$\EHrightnat(1_{1_\star},\p)$}};
\node at (4, 3) {\emph{$\EHleftnat(\q,1_{1_\star})$}};
\draw[-] (N0a) -- node[above]{$1_{1_{1_\star}}$} (N1a);
\draw[-] (N1a) -- node[above]{$1_{1_{1_\star}}$} (N2a);
\draw[-] (N0b) -- node[above]{$1_{1_{1_\star}}$} (N1b);
\draw[-] (N1b) -- node[above]{$1_{1_{1_\star}}$} (N2b);
\draw[-] (N0c) -- node[below]{$1_{1_{1_\star}}$} (N1c);
\draw[-] (N1c) -- node[below]{$1_{1_{1_\star}}$} (N2c);
\draw[-] (N0a) -- node[left]{\emph{$\whiskerL(1_{1_\star},\p)$}} (N0b);
\draw[-] (N0b) -- node[left]{\emph{$\whiskerR(\q,1_{1_\star})$}} (N0c);
\draw[-] (N2a) -- node[right]{\emph{$\whiskerR(\p,1_{1_\star})$}} (N2b);
\draw[-] (N2b) -- node[right]{\emph{$\whiskerL(1_{1_\star},\q)$}} (N2c);
\end{tikzpicture}
\end{center}
We have already established a relationship between these in a previous section:
\emph{
\begin{align*}
& \EHrightnat(1_{1_\star},\p) = \concatleftnat(\p) \, \concathor \; \concatrightnat(\p)^\invhor \\
& \EHleftnat(\q,1_{1_\star}) = \concatrightnat(\q) \, \concathor \; \concatleftnat(\q)^\invhor
\end{align*}}%
Thus, we can abstractly summarize the entire situation in the following lemma:

\begin{lemma}
Assume points $a, b, c : A$, 1-paths $p,q,r : a = b$, $u,v,w : b = c$, and 2-paths $\alpha : p \ct 1_b = 1_a \ct q$, $\beta : r \ct 1_b = 1_a \ct q$, $\gamma : u \ct 1_c = 1_b \ct v$, $\delta : w \ct 1_c = 1_b \ct v$ as in the diagram below:

\begin{center}
\begin{tikzpicture}
\node (N0a) at (0,6) {$a$};
\node (N1a) at (2,6) {$a$};
\node (N2a) at (4,6) {$a$};
\node (N0b) at (0,4) {$b$};
\node (N1b) at (2,4) {$b$};
\node (N2b) at (4,4) {$b$};
\node (N0c) at (0,2) {$c$};
\node (N1c) at (2,2) {$c$};
\node (N2c) at (4,2) {$c$};
\node at (1, 5) {$\alpha$};
\node at (3, 5) {$\beta^\invhor$};
\node at (1, 3) {$\gamma$};
\node at (3, 3) {$\delta^\invhor$};
\draw[-] (N0a) -- node[above]{$1_a$} (N1a);
\draw[-] (N1a) -- node[above]{$1_a$} (N2a);
\draw[-] (N0b) -- node[above]{$1_b$} (N1b);
\draw[-] (N1b) -- node[above]{$1_b$} (N2b);
\draw[-] (N0c) -- node[below]{$1_c$} (N1c);
\draw[-] (N1c) -- node[below]{$1_c$} (N2c);
\draw[-] (N0a) -- node[left]{$p$} (N0b);
\draw[-] (N0b) -- node[left]{$u$} (N0c);
\draw[-] (N1a) -- node[left]{$q$} (N1b);
\draw[-] (N1b) -- node[left]{$v$} (N1c);
\draw[-] (N2a) -- node[right]{$r$} (N2b);
\draw[-] (N2b) -- node[right]{$w$} (N2c);
\end{tikzpicture}
\end{center}
Furthermore, assume 2-paths $\theta : p \ct 1_b = 1_a \ct r$ and $\phi : u \ct 1_c = 1_b \ct w$ as in the diagram below:
\begin{center}
\begin{tikzpicture}
\node (N0a) at (0,6) {$a$};
\node (N1a) at (2,6) {$a$};
\node (N2a) at (4,6) {$a$};
\node (N0b) at (0,4) {$b$};
\node (N1b) at (2,4) {$b$};
\node (N2b) at (4,4) {$b$};
\node (N0c) at (0,2) {$c$};
\node (N1c) at (2,2) {$c$};
\node (N2c) at (4,2) {$c$};
\node at (2, 5) {$\theta$};
\node at (2, 3) {$\phi$};
\draw[-] (N0a) -- node[above]{$1_a$} (N1a);
\draw[-] (N1a) -- node[above]{$1_a$} (N2a);
\draw[-] (N0b) -- node[above]{$1_b$} (N1b);
\draw[-] (N1b) -- node[above]{$1_b$} (N2b);
\draw[-] (N0c) -- node[below]{$1_c$} (N1c);
\draw[-] (N1c) -- node[below]{$1_c$} (N2c);
\draw[-] (N0a) -- node[left]{$p$} (N0b);
\draw[-] (N0b) -- node[left]{$u$} (N0c);
\draw[-] (N2a) -- node[right]{$r$} (N2b);
\draw[-] (N2b) -- node[right]{$w$} (N2c);
\end{tikzpicture}
\end{center}
If $\theta = \alpha \, \concathor \, \beta^\invhor$ and $\phi = \gamma \, \concathor \, \delta^\invhor$, then the following triangle commutes:
\begin{center}
\begin{tikzpicture}
\node (N0) at (0,0) {$p \ct u$};
\node (N1) at (1.5,2) {$q \ct v$};
\node (N2) at (3,0) {$r \ct w$};
\draw[->] (N1) -- node[left]{\scriptsize $\big(\I(\alpha) \ct\ct \I(\gamma)\big)^{-1}$} (N0);
\draw[->] (N1) -- node[right]{\scriptsize $\big(\I(\beta) \ct\ct \I(\delta)\big)^{-1}$} (N2);
\draw[->] (N0) -- node[below]{\scriptsize $\I(\theta \concatvert \phi)$} (N2);
\end{tikzpicture}
\end{center}
\end{lemma}
To prove the above lemma, we first perform path induction on the two hypotheses, thereby eliminating the 2-paths $\theta$ and $\phi$. Next we reformulate the goal so that we can get rid of the remaining 2-paths by induction - \emph{given points $a, b, c : A$, 1-paths $p,q,r : a = b$, $u,v,w : b = c$, and 2-paths $\alpha : p = q$, $\beta : r = q$, $\gamma : u = v$, $\delta : w = v$, the following triangle commutes}:
\begin{center}
\begin{tikzpicture}
\node (N0) at (0,0) {$p \ct u$};
\node (N1) at (4.5,2.5) {$q \ct v$};
\node (N2) at (9,0) {$r \ct w$};
\draw[->] (N1) -- node[left]{\scriptsize $\big(\I(\I^{-1}(\alpha)) \ct\ct \I(\I^{-1}(\gamma))\big)^{-1}$} (N0);
\draw[->] (N1) -- node[right]{\scriptsize \;\; $\big(\I(\I^{-1}(\beta)) \ct\ct \I(\I^{-1}(\delta))\big)^{-1}$} (N2);
\draw[->] (N0) -- node[below]{\scriptsize $\I\Big(\big(\I^{-1}(\alpha) \concathor (\I^{-1}(\beta))^\invhor\big) \concatvert \big(\I^{-1}(\gamma) \concathor (\I^{-1}(\delta))^\invhor\big)\Big)$} (N2);
\end{tikzpicture}
\end{center}
Now we perform path induction on $\alpha,\beta,\gamma,\delta$, which also eliminates the 1-paths $p,r,v,w$. The only two remaining 1-paths are $p : a = b$ and $u : b =c$, and these beg for further path induction after which there is nothing left to do.

The lemma now immediately implies the commutativity of the upper triangle. In fact, it also implies the commutativity of the lower triangle - the four commuting squares $\concatleftnat(\p), \concatrightnat(\p), \concatleftnat(\q), \concatrightnat(\q)$ can now be put together as follows:

\begin{center}
\begin{tikzpicture}
\node (N0a) at (0,6) {$1_{1_\star} \ct 1_{1_\star}$};
\node (N1a) at (4,6) {$1_{1_\star}$};
\node (N2a) at (8,6) {$1_{1_\star} \ct 1_{1_\star}$};
\node (N0b) at (0,4) {$1_{1_\star} \ct 1_{1_\star}$};
\node (N1b) at (4,4) {$1_{1_\star}$};
\node (N2b) at (8,4) {$1_{1_\star} \ct 1_{1_\star}$};
\node (N0c) at (0,2) {$1_{1_\star} \ct 1_{1_\star}$};
\node (N1c) at (4,2) {$1_{1_\star}$};
\node (N2c) at (8,2) {$1_{1_\star} \ct 1_{1_\star}$};
\node at (2, 5) {\emph{$\concatrightnat(\q)$}};
\node at (6, 5) {\emph{$\concatleftnat(\q)^\invhor$}};
\node at (2, 3) {\emph{$\concatleftnat(\p)$}};
\node at (6, 3) {\emph{$\concatrightnat(\p)^\invhor$}};
\draw[-] (N0a) -- node[above]{$1_{1_{1_\star}}$} (N1a);
\draw[-] (N1a) -- node[above]{$1_{1_{1_\star}}$} (N2a);
\draw[-] (N0b) -- node[above]{$1_{1_{1_\star}}$} (N1b);
\draw[-] (N1b) -- node[above]{$1_{1_{1_\star}}$} (N2b);
\draw[-] (N0c) -- node[below]{$1_{1_{1_\star}}$} (N1c);
\draw[-] (N1c) -- node[below]{$1_{1_{1_\star}}$} (N2c);
\draw[-] (N0a) -- node[left]{\emph{$\whiskerR(\q,1_{1_\star})$}} (N0b);
\draw[-] (N0b) -- node[left]{\emph{$\whiskerL(1_{1_\star},\p)$}} (N0c);
\draw[-] (N1a) -- node[left]{\emph{$\q$}} (N1b);
\draw[-] (N1b) -- node[left]{\emph{$\p$}} (N1c);
\draw[-] (N2a) -- node[right]{\emph{$\whiskerL(1_{1_\star},\q)$}} (N2b);
\draw[-] (N2b) -- node[right]{\emph{$\whiskerR(\p,1_{1_\star})$}} (N2c);
\end{tikzpicture}
\end{center}
The two commuting squares $\EHleftnat(1_{1_\star},\p)$ and $\EHrightnat(\q,1_{1_\star})$ can be stacked as in the diagram below:
\begin{center}
\begin{tikzpicture}
\node (N0a) at (0,6) {$1_{1_\star} \ct 1_{1_\star}$};
\node (N1a) at (4,6) {$1_{1_\star}$};
\node (N2a) at (8,6) {$1_{1_\star} \ct 1_{1_\star}$};
\node (N0b) at (0,4) {$1_{1_\star} \ct 1_{1_\star}$};
\node (N1b) at (4,4) {$1_{1_\star}$};
\node (N2b) at (8,4) {$1_{1_\star} \ct 1_{1_\star}$};
\node (N0c) at (0,2) {$1_{1_\star} \ct 1_{1_\star}$};
\node (N1c) at (4,2) {$1_{1_\star}$};
\node (N2c) at (8,2) {$1_{1_\star} \ct 1_{1_\star}$};
\node at (4, 5) {\emph{$\EHleftnat(\q,1_{1_\star})$}};
\node at (4, 3) {\emph{$\EHrightnat(1_{1_\star},\p)$}};
\draw[-] (N0a) -- node[above]{$1_{1_{1_\star}}$} (N1a);
\draw[-] (N1a) -- node[above]{$1_{1_{1_\star}}$} (N2a);
\draw[-] (N0b) -- node[above]{$1_{1_{1_\star}}$} (N1b);
\draw[-] (N1b) -- node[above]{$1_{1_{1_\star}}$} (N2b);
\draw[-] (N0c) -- node[below]{$1_{1_{1_\star}}$} (N1c);
\draw[-] (N1c) -- node[below]{$1_{1_{1_\star}}$} (N2c);
\draw[-] (N0a) -- node[left]{\emph{$\whiskerR(\q,1_{1_\star})$}} (N0b);
\draw[-] (N0b) -- node[left]{\emph{$\whiskerL(1_{1_\star},\p)$}} (N0c);
\draw[-] (N2a) -- node[right]{\emph{$\whiskerL(1_{1_\star},\q)$}} (N2b);
\draw[-] (N2b) -- node[right]{\emph{$\whiskerR(\p,1_{1_\star})$}} (N2c);
\end{tikzpicture}
\end{center}
And as before, we have
\emph{
\begin{align*}
& \EHleftnat(\q,1_{1_\star}) = \concatrightnat(\q) \, \concathor \; \concatleftnat(\q)^\invhor \\
& \EHrightnat(1_{1_\star},\p) = \concatleftnat(\p) \, \concathor \; \concatrightnat(\p)^\invhor
\end{align*}}%
The same lemma thus implies the commutativity of the following triangle:
\begin{center}
\begin{tikzpicture}
\node (N0b) at (0,0) {\emph{$\whiskerR(\q,1_{1_\star}) \ct \whiskerL(1_{1_\star},\p)$}};
\node (N1b) at (5.5,2.5) {\emph{$\q \ct \p$}};
\node (N2b) at (11,0) {\emph{$\whiskerL(1_{1_\star},q) \ct \whiskerR(p,1_{1_\star})$}};
\node at (-0.2,1.4) {\scriptsize \emph{$\Big(\I\big(\concatrightnat(\q)\big) \ct\ct \I\big(\concatleftnat(\p)\big)\Big)^{-1}$}};
\node at (11.5,1.4) {\scriptsize \emph{$\Big(\I\big(\concatleftnat(\q)\big) \ct\ct \I\big(\concatrightnat(\p)\big)\Big)^{-1}$}};
\draw[->] (N1b) -- node[left]{} (N0b);
\draw[->] (N1b) -- node[right]{} (N2b);
\draw[->] (N0b) -- node[below]{\scriptsize \emph{$\I\Big(\EHleftnat(\q,1_{1_\star}) \concatvert \EHrightnat(1_{1_\star},\p)\Big)$}} (N2b);
\end{tikzpicture}
\end{center}
Flipping the triangle along the horizontal axis yields precisely the desired lower triangle, and we are done.


\paragraph{Step 3}
It remains to somehow combine the two commuting triangles on the top and bottom with the commuting square in the middle. We cannot literally paste them together because, as we recall, the vertices of the square contain an extra reflexivity path originating from the Eckmann-Hilton term. But this is not a problem because we have a suitable generalization up our sleeve:

\begin{lemma}
Assume points $a, b : A$, 1-paths $p,q,u,v,x,y : a = b$, and 2-paths $\alpha : u = p$, $\beta : v = p$, $\gamma : x = q$, $\delta : y = q$, $\theta : u = x$, $\phi : y = v$ as in the diagram below:
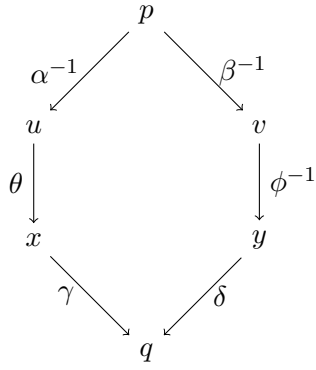
\begin{center}
\begin{tikzpicture}
\node (N0) at (1.5,4.5) {$p$};
\node (N1) at (0,3) {$u$};
\node (N2) at (0,1.5) {$x$};
\node (N3) at (1.5,0) {$q$};
\node (N4) at (3,1.5) {$y$};
\node (N5) at (3,3) {$v$};
\draw[->] (N0) -- node[left]{$\alpha^{-1}$} (N1);
\draw[->] (N1) -- node[left]{$\theta$} (N2);
\draw[->] (N2) -- node[left]{$\gamma$\,} (N3);
\draw[->] (N4) -- node[right]{$\delta$} (N3);
\draw[->] (N5) -- node[right]{$\phi^{-1}$} (N4);
\draw[->] (N0) -- node[right]{\,$\beta^{-1}$} (N5);
\end{tikzpicture}
\end{center}
Assume further 2-paths $\eta : u \ct 1_b = 1_a \ct v$ and $\varepsilon : x \ct 1_b = 1_a \ct y$. Then the hexagon above commutes provided the triangles and square below do:
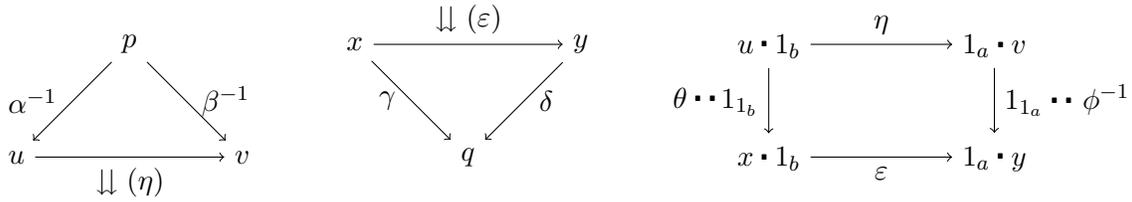
\begin{center}
\begin{tikzpicture}
\node (N0a) at (1.5,1.5) {$p$};
\node (N1a) at (0,0) {$u$};
\node (N5a) at (3,0) {$v$};
\node (N0b) at (4.5,1.5) {$x$};
\node (N1b) at (7.5,1.5) {$y$};
\node (N5b) at (6,0) {$q$};
\draw[->] (N0a) -- node[left]{$\alpha^{-1}$\,} (N1a);
\draw[->] (N0a) -- node[right]{\,$\beta^{-1}$} (N5a);
\draw[->] (N1a) -- node[below]{$\I(\eta)$} (N5a);
\draw[->] (N0b) -- node[above]{$\I(\varepsilon)$} (N1b);
\draw[->] (N0b) -- node[left]{$\gamma$\,} (N5b);
\draw[->] (N1b) -- node[right]{\,$\delta$} (N5b);
\node (N1c) at (10,1.5) {$u \ct 1_b$};
\node (N2c) at (10,0) {$x \ct 1_b$};
\node (N4c) at (13,0) {$1_a \ct y$};
\node (N5c) at (13,1.5) {$1_a \ct v$};
\draw[->] (N1c) -- node[left]{$\theta \ctt 1_{1_b}$} (N2c);
\draw[->] (N5c) -- node[right]{$1_{1_a} \ctt \, \phi^{-1}$} (N4c);
\draw[->] (N1c) -- node[above]{$\eta$} (N5c);
\draw[->] (N2c) -- node[below]{$\varepsilon$} (N4c);
\end{tikzpicture}
\end{center}
\end{lemma}
To prove this lemma, we first perform path induction on $\theta, \phi, \beta, \delta$. This eliminates the 1-paths $p, q, v, y$. The commutative square hypothesis now becomes equivalent to $\eta = \varepsilon$, so we can perform induction and get rid of $\varepsilon$. The commutative triangle hypotheses now become equivalent to $\I(\eta) = \alpha$ and $\I(\eta) = \gamma$, so we can perform induction and get rid of $\alpha$ and $\gamma$. Among the 2-paths this only leaves $\eta : u \ct 1_b = 1_a \ct y$, which is equivalent to $\eta : u = y$, so we can perform induction to get rid of $y$. The sole remaining 1-path is $u : a = b$, and we perform one last induction on it. We have managed to reduce everything in sight to reflexivity and thus made the hexagon trivially commute.

The conclusion of the above lemma clearly implies syllepsis, and the previous two steps show that the hypotheses hold, so $\checkmark$.

\bibliographystyle{plainnat}
\bibliography{references} 

\end{document}